# Spinors and Rodrigues representations associated with orthogonal polynomials

Z. Bakhshi*

Department of Physics, Faculty of Basic Sciences, Shahed University, Tehran, Iran.

August 2, 2018

## Abstract

An effective approach is presented to produce Schrödinger-like equation for the spinor components from Dirac equation. Considering electrostatic potential as a constant value yields a second-order differential equation that is comparable with the well-known solvable models in the non-relativistic quantum mechanics for the certain bound state energy spectrum and the well-known potentials. By this comparison, the gage field potential and the relativistic energy can be written by the non-relativistic models and the spinors will be related to the orthogonal polynomials. It has also shown that the upper spinors wave functions based on the orthogonal polynomials can be given in terms of the Rodrigues representations. Association with the Rodrigues representations of orthogonal polynomials have also been investigated in the lower spinor components, since they are related to the upper spinor components according to first-order differential equation that is attained from Dirac equation.

*(E-mail: z.bakhshi@shahed.ac.ir)

# 1 Introduction

In recent years, there has been a developing interest in search for exactly solvable systems in non-relativistic and relativistic quantum mechanics. Exactly solvable meaning the eigenvalues and the eigenfunctions of the Hamiltonian operator of the physical system can be derived analytically in closed form. Solvable models are noteworthy because understanding of physics can only be brought with such solutions. Moreover, exact solutions are valuable tools for testing and improving numerical methods introduced to solve problems physically more interesting [1]. Since, relativistic extensions of the exact solvable potentials are very useful to study the relativistic effects, various methods were employed to obtain the exact solution of the problem. Point canonical transformation [2-4], dynamical group [5,6], factorization method [7], supersymetric quantum mechanics and shape invariance [8-10] are methods among many which were used in the search for exact solutions of wave function. Also, there are a lot of investigations that show how methods used to obtain analytical solutions of the Schrödinger equation can be extended to Dirac case [11-15].

Alhaidari [11-13] applies a unitary transformation to Dirac equation such that the resulting second-order differential equation becomes Schrödinger-like equation so that comparison with the well-known non-relativistic problems is transparent. If the electrostatic potential is assumed as a constant value, the second-order differential equation can be constituted for upper component by eliminating lower component, without applying a general local unitary transformation that eliminates the first-order derivative such as what Alhaidary has considered.

In this method, by assuming electrostatic potential as a constant value, the second-order differential equation can be compared with the well-known solvable Schrödinger equation in the non-relativistic quantum models. The wave functions in Schrödinger equation for the well-known potentials have been obtained on the orthogonal polynomials, such as Jacobi, Generalized Lauegrre and Hermite polynomials and the energy eigenvalues spectrum can be accessible for each case. By comparing the second-order differential equation that has been obtained from Dirac equation with Schrödinger equation for the well-known potential such as Scarff-II, Pöshel-Teller, Mörse, 3D-oscillator and shift-oscillator potentials, the gage field potential can be written based on the well-known superpotentials that are related to the mentioned potentials. Therefore, the second-order differential is transformed to the solvable models with the exact solutions, it means that the relativistic energy eigenvalues can be gotten based on the non-relativistic models, also the spinors will be related to the orthogonal polynomials according to the non-relativistic models. Then, Rodriguues representations and the differential equation of them are calculated for orthogonal polynomials. Moreover, the second-order differential equation also can be considered as a product of two first-order differential operators and the spinor wave function related to the differential equation that is expressed in terms of Rodrigues representations that is related to the orthogonal polynomials. Therefore, the solution of second-order differential equation can be considered with the determined relativistic energy and association with Rodriguues representations can be gotten for each orthogonal polynomials.

This paper is organized as followed: In section 2, by using the point canonical transformations, the second-order differential equation is constituted with the gauge field potential and the energy spectrum that will be introduced based on the non-relativistic models. Then, the association of Rodriguues representation with orthogonal polynomials are shown in the sections 3, 4 and 5 for Jacobi, Generalized Lauegrre and Hermite polynomials, respectively. In each section, all of the gage field potential are considered that they can constituted the solvable models with the certain energy eigenvalues, for each orthogonal polynomials. Therefore, in the each section, Rodriguues representations of the orthogonal polynomials have been calculated for some gage field potentials. In section 6, the paper ends with a brief conclusion.

## 2 The three-dimensional Dirac equation for a free structure

Particle of spin $\frac{1}{2}$ reads $(i\hbar\gamma^\mu\partial_\mu - mc)\Phi = 0$, where $m$ is the rest mass of the particle, $c$ is the speed of light, and $\Phi$ is a four-component wave function. The four matrices $\{\gamma^\mu\}_{\mu=0}^3$ are given the following standard representation [16]

$$\gamma^0 = \begin{pmatrix} I & 0 \\ 0 & -I \end{pmatrix}, \qquad \vec{\gamma} = \begin{pmatrix} 0 & \vec{\sigma} \\ -\vec{\sigma} & 0 \end{pmatrix}, \qquad (2.1)$$

where $I$ is the $2 \times 2$ unit matrix and $\sigma$ are the usual $2 \times 2$ Pauli spin matrices. In atomic units ($m = e = \hbar = 1$), Dirac equation reads $(i\gamma^\mu\partial_\mu - \alpha^{-1})\Phi = 0$, where $\alpha = \frac{\hbar}{mc} = \frac{1}{c}$ is the Compton wavelength of the particle. In the presence of the electromagnetic potential, $A_\mu = (A_0, c\vec{A})$, gauge invariant coupling to the charged spinor is accomplished by the minimal substitution $\partial_\mu \to \partial_\mu + i\alpha A_\mu$, which transforms free Dirac equation into

$$[i\gamma^\mu(\partial_\mu + i\alpha A_\mu) - \alpha^{-1}]\Phi = 0. \qquad (2.2)$$

For time independent potential, Eq.(2.2) gives the following matrix representation of Dirac Hamiltonian (in units of $mc^2 = \alpha^2$) [14]

$$H = \begin{pmatrix} \alpha^2 A_0 + 1 & -i\alpha\vec{\sigma}.\vec{\nabla} + \alpha\vec{\sigma}.\vec{A} \\ -i\alpha\vec{\sigma}.\vec{\nabla} + \alpha\vec{\sigma}.\vec{A} & \alpha^2 A_0 - 1 \end{pmatrix}. \qquad (2.3)$$

Taking into consideration gauge invariance, the form of electromagnetic potential for static charge distribution with spherical symmetry is

$$(A_0, \vec{A}) = (v(r), \hat{r}\omega(r)), \qquad (2.4)$$

where $\hat{r}$ is radial unit vector, $v(r)$ and $\omega(r)$ are electrostatic potential and gauge field potential, respectively. By substituting the two off-diagonal term $\alpha\vec{\sigma}.\vec{A}$ by $\pm i\alpha\vec{\sigma}.\vec{A}$ in (2.3), the

Hamiltonian leads to the following two component radial Dirac equation [17]

$$\begin{pmatrix} \alpha^2 v(r)+1 & \alpha(\frac{k}{r}+\omega(r)-\frac{d}{dr}) \\ \alpha(\frac{k}{r}+\omega(r)+\frac{d}{dr}) & \alpha^2 v(r)-1 \end{pmatrix} \begin{pmatrix} \varphi(r) \\ \theta(r) \end{pmatrix} = \varepsilon \begin{pmatrix} \varphi(r) \\ \theta(r) \end{pmatrix}, \qquad (2.5)$$

where $\varepsilon$ is the relativistic energy eigenvalues and $k$ is the spin-orbit coupling parameter defined as $k = \pm(j+\frac{1}{2})$ for $l = j \pm \frac{1}{2}$. Eq.(2.5) gives two coupled first-order differential equations for the radial spinor components. By eliminating lower spinor component and by assuming that the electrostatic potential $v(r)$ to be a constant value $\eta$, the second-order differential equation can be gotten for upper spinor wave function as

$$-\frac{d^2\varphi}{dr^2} + [(\omega(r)+\frac{k}{r})^2 - (\frac{d\omega}{dr}-\frac{k}{r^2}) - (\frac{(\alpha^2\eta-\varepsilon)^2-1}{\alpha^2})]\varphi(r) = 0. \qquad (2.6)$$

Eq.(2.5) also gives the lower spinor component in terms of the upper component as followed

$$\theta(r) = [\alpha\eta - (\frac{\varepsilon+1}{\alpha})]^{-1} \left\{ [\omega(r)+\frac{k}{r}]\varphi(r) + \frac{d\varphi}{dr} \right\}. \qquad (2.7)$$

By comparing Eq.(2.6) with the solvable Schrödinger equation in the non-relativistic models, the relation can be considered between the well-known potential in the non-relativistic quantum models and the gage field potential in the relativistic system as $V_m(r) = (\omega(r)+\frac{k}{r})^2 - (\frac{d\omega}{dr}-\frac{k}{r^2})$. Also, non-relativistic energy eigenvalues can be related to the relativistic energy eigenvalues as $E = \frac{(\alpha^2\eta-\varepsilon)^2-1}{\alpha^2}$. So, the gauge field potential and the relativistic energy that due to solvability of Dirac equation based on the non relativistic quantum mechanics are easily available.

## 3 Association of Rodruiges representation with Jacobi polynomials

Let us consider the gauge field potentials where their wave functions are related to Jacobi polynomials such as Pöschl-Teller potential $\omega^{(1)}(r) = -A\coth r + \frac{B}{\sinh r} - \frac{k}{r}$ and Scarf-II potential $\omega^{(2)}(r) = -A\tanh r - \frac{B}{\cosh r} - \frac{k}{r}$ where $A$ and $B$ are real parameters. For each potential, respectively, Eq.(2.6) gives the following second-order differential equations for upper spinor component

$$-\frac{d^2\varphi_{n,m}^{(1)}(r)}{dr^2} + \left[A^2 + \frac{(B^2 - A^2 + A)}{\sinh^2 r} + \frac{(B - 2AB)\cosh r}{\sinh^2 r}\right]\varphi_{n,m}^{(1)}(r) = \left[\frac{(\alpha^2\eta - \varepsilon)^2 - 1}{\alpha^2}\right]\varphi_{n,m}^{(1)}(r) \quad (3.1)$$

$$-\frac{d^2\varphi_{n,m}^{(2)}(r)}{dr^2} + \left[A^2 + \frac{(B^2 - A^2 + A)}{\cosh^2 r} + \frac{(2AB - B)\sinh r}{\cosh^2 r}\right]\varphi_{n,m}^{(2)}(r) = \left[\frac{(\alpha^2\eta - \varepsilon)^2 - 1}{\alpha^2}\right]\varphi_{n,m}^{(2)}(r) \quad (3.2)$$

where $A = \frac{\lambda+\gamma+2m-1}{2}$ and $B = \frac{\gamma-\lambda}{2}$ such that $\lambda, \gamma > -1$ in Eq.(3.1) and $A = m + \lambda - \frac{1}{2}$ and $B = \frac{\gamma}{2}$ such that $\lambda > -1$ and $-\infty > \gamma > +\infty$ in Eq.(3.2). There will be the well-known non-relativistic energy spectrum as $E_{n,m}^{(1)} = (\lambda + \gamma + n + m)(m - n - 1)$ and $E_{n,m}^{(2)} = (2\lambda + n + m)(n - m + 1)$ for Pöschl-Teller potential and Scarf-II potential, respectively. As mentioned before, they can be used to calculate relativistic energy spectrum of Dirac equation as the following forms

$$\varepsilon_n^{(1)} = \alpha^2\eta \mp \left[-\alpha^2(\lambda + \gamma + n + m)(n - m + 1) + 1\right]^{\frac{1}{2}}, \quad (3.3)$$

$$\varepsilon_n^{(2)} = \alpha^2\eta \mp \left[-\alpha^2(2\lambda + n + m)(n - m + 1) + 1\right]^{\frac{1}{2}}. \quad (3.4)$$

The bound states wave functions of the non-relativistic problem [18] are mapped into the following upper spinor components wave functions

$$\varphi_{n,m}^{(1)}(x) \propto (x-1)^{\frac{2\lambda+2m-1}{4}} \times (x+1)^{\frac{2\gamma+2m-1}{4}} P_n^{(\lambda+m-1,\gamma+m-1)}(x), \quad (3.5)$$

$$\varphi_{n,m}^{(2)}(x) \propto \left(1+x^2\right)^{-\frac{1}{2}} \times \exp\left(\frac{\gamma}{2}\tanh^{-1} x\right) P_n^{(i\frac{\gamma}{2}+m+\lambda-\frac{1}{4},-i\frac{\gamma}{2}+m+\lambda-\frac{1}{4})}(x), \quad (3.6)$$

where $P_n^{(\mu,\nu)}(x)$ is Jacobi polynomial with $\mu, \nu > -1$, and $x = \cosh r$, $\mu = \lambda + m - 1$ and $\nu = \gamma + m - 1$ in Eq.(3.5) and $x = \sinh r$, $\mu + i\frac{\gamma}{2} + m + \lambda - \frac{1}{4}$ and $\nu = -i\frac{\gamma}{2} + m + \lambda - \frac{1}{4}$ in Eq.(3.6). By substituting upper spinor component (3.5) and (3.6) into Eq.(2.7) and using recursion properties of Jacobi polynomials, lower spinor components are given as

$$\theta_{n,m}^{(1)}(x) \propto \left(\frac{1}{\alpha} \mp \left[\frac{1}{\alpha^2} - n^2 - n(\lambda + \gamma + 2m - 1)\right]^{\frac{1}{2}}\right)^{-1}$$

$$\times \left( \left[ nx - (\frac{\lambda - \gamma}{2n + 2m + \lambda + \gamma - 2}) \right] (x^2 - 1)^{-\frac{1}{2}} \varphi_{n,m}^{(1)}(x) \right.$$
$$\left. - \left[ \frac{2(n + m + \lambda - 1)(n + m + \gamma - 1)}{2n + 2m + \lambda + \gamma - 2} \right] (x^2 - 1)^{-\frac{1}{2}} \varphi_{n-1,m}^{(1)}(x) \right), \quad (3.7)$$

$$\theta_{n,m}^{(2)}(x) \propto \left( \frac{1}{\alpha} \mp \left[ \frac{1}{\alpha^2} + (n - m + 1)(m - n + 2\lambda - 2) \right]^{\frac{1}{2}} \right)^{-1}$$
$$\times \left( \left[ -(n + \lambda + \frac{3}{4})x + \frac{\gamma}{2} \left( \frac{2n - 3m - \lambda + 3}{-n + 2m - 2} \right) \right] (x^2 + 1)^{-\frac{1}{2}} \right.$$
$$\left. + \frac{\gamma}{2}(1 - x)^{-1}(x^2 + 1)^{\frac{1}{2}} \varphi_{n,m}^{(2)}(x) - \left[ \frac{i(-n + 2m + \lambda - 2)^2 + \frac{\gamma^2}{8}}{-n + 2m + \lambda + -2} \right] (x^2 + 1)^{-\frac{1}{2}} \varphi_{n-1,m}^{(2)}(x) \right) \quad (3.8)$$

Raising and lowering operators $B_{\pm}(m) = \pm \frac{d}{dr} + W_m(x(r))$, where the superpotential $W_m(x(r))$ satisfies in the Riccati equation $V_m = W_m^2 \pm W_m'$, can be written as following forms according to Pöschl-Teller and Scarf-II potentials, respectively,

$$B_+^{(1)} = \frac{d}{dr} + [-A \coth r + \frac{B}{\sinh r}], \quad B_-^{(1)} = -\frac{d}{dr} + [-A \coth r + \frac{B}{\sinh r}]. \quad (3.9)$$

$$B_+^{(2)} = \frac{d}{dr} + [-A \tanh r - \frac{B}{\cosh r}], \quad B_-^{(2)} = -\frac{d}{dr} + [-A \tanh r + \frac{B}{\cosh r}]. \quad (3.10)$$

It is obvious that the second-order differential equations can always be considered in a factorization form as a product of a pair of linear differential operators (3.9) and (3.10). Therefore,

$$B_+^{(1),(2)}(m) B_-^{(1),(2)}(m) \varphi_{n,m}^{(1),(2)}(r) = E^{(1),(2)}(n, m) \varphi_{n,m}^{(1),(2)}(r), \quad (3.11)$$

$$B_-^{(1),(2)}(m) B_+^{(1),(2)}(m) \varphi_{n,m-1}^{(1),(2)}(r) = E^{(1),(2)}(n, m) \varphi_{n,m-1}^{(1),(2)}(r). \quad (3.12)$$

In the above equations, for a given $n$, the operators $B_+(m)$ raises the index $m$ while the operator $B_-(m)$ lower it. We can also obtain the highest state $\varphi_{n,n}$ by solving the first-order differential equation $B_+(n+1)\varphi_{n,n}(r) = 0$ because the non-relativistic energy spectrum $E(n, m)$ vanishes for $m = n + 1$ [10]. Since, by introducing a new function as $\varphi_{n,m}(x) = A^{\frac{1}{4}}(x) W^{\frac{1}{2}}(x) \psi_{n,m}(x)$ and changing the variable $\frac{dx}{dr} = \sqrt{A(x)}$, Schrödinger equation (2.6) has been obtained from the general form of associated with second-order differential equation in terms of master function $A(x)$ and the wave function $W(x)$ as follows [19-21]

$$A(x)\ddot{\psi}_{n,m}(x) + \frac{(A(x)W(x))'}{W(x)}\dot{\psi}_{n,m}(x) + \left[ -\frac{1}{2}(n^2+n-m^2)\ddot{A}(x) + (m-n)\left(\frac{A(x)\dot{W}(x)}{W(x)}\right)' \right.$$
$$\left. - \frac{m^2}{4}\frac{\dot{A}(x)^2}{A(x)} - \frac{m}{2}\frac{\dot{A}(x)\dot{W}(x)}{W(x)} \right]\psi_{n,m}(x) = 0, \quad (3.13)$$

where $\psi_{n,m}(x)$ is Rodruiges representation of the orthogonal polynomials in Eq. (3.13). For a positive integer $n$, $\psi_{n,m}(x)$ as Rodruiges representationis is given by

$$\psi_{n,m}(x) = (-1)^m A^{\frac{m}{2}}(x) \left(\frac{d}{dx}\right)^m \psi_n(x), \quad m = 0, 1, 2, .., n \quad (3.14)$$

where

$$\psi_n(x) = \frac{N}{W(x)} \left(\frac{d}{dx}\right)^n (A^n(x)W(x)), \quad (3.15)$$

with $N$ which is normalization constant. So, for each case of the gage field potentials, the Rodruiges representations of upper and lower spinors and differential equations associated with them are available. Furthermore, it can be shown that they are clear examples of connection between Jocobi polynomials and Rodruiges representations in Dirac equation. Since, the wave functions $\psi_{n,m}(x)$ are related to the upper spinors as the following forms

$$\psi_{n,m}^{(1)}(x) = (x-1)^{-(\frac{\lambda}{2}+\frac{1}{4})}(x+1)^{-(\frac{\gamma}{2}+\frac{1}{4})} \varphi_{n,m}^{(1)}(x), \quad (3.16)$$

for, $A^{(1)}(x) = x^2 - 1$ and $W^{(1)}(x) = (x-1)^\lambda (x+1)^\gamma$ in Pöschl-Teller potential and,

$$\psi_{n,m}^{(2)}(x) = \left(x^2+1\right)^{-(\frac{\lambda}{2}+\frac{1}{4})} \exp\left(-\frac{\gamma}{2}\tan^{-1}(x)\right) \varphi_{n,m}^{(2)}(x), \quad (3.17)$$

when, $A^{(2)}(x) = x^2 + 1$ and $W^{(2)}(x) = (x^2+1)^\lambda \exp(\gamma \tan^{-1}(x))$ in Scarf-II potential. Since the lower spinors can be connected to the upper spinors according to the Eqs.(3.7) and (3.8), therefore, Rodruiges representations of upper spinors also can be associated with lower spinors. If the wave function $\Theta_{n,m}(x)$ is introduced for lower spinor $\theta_{n,m}(x)$, it can be written based on Rodruiges representations $\psi_{n,m}(x)$. The wave function $\Theta_{n,m}(x)$ that is connected to the lower spinors $\theta_{n,m}(x)$ can be written as follows for Pöschl-Teller and Scarf-II potentials, proportionately,

$$\Theta^{(1)}_{n,m}(x) \propto \left(\frac{1}{\alpha} \mp \left[\frac{1}{\alpha^2} - n^2 - n(\lambda + \gamma + 2m - 1)\right]^{\frac{1}{2}}\right)^{-1}$$
$$\times \left(\left[nx - \left(\frac{\lambda - \gamma}{2n + 2m + \lambda + \gamma - 2}\right)\right](x^2 - 1)^{-\frac{1}{2}} \psi^{(1)}_{n,m}(x)\right.$$
$$\left. - \left[\frac{2(n + m + \lambda - 1)(n + m + \gamma - 1)}{2n + 2m + \lambda + \gamma - 2}\right](x^2 - 1)^{-\frac{1}{2}}\psi^{(1)}_{n-1,m}(x)\right) \quad (3.18)$$

and

$$\Theta^{(2)}_{n,m}(x) \propto \left(\frac{1}{\alpha} \mp \left[\frac{1}{\alpha^2} + (n - m + 1)(m - n + 2\lambda - 2)\right]^{\frac{1}{2}}\right)^{-1}$$
$$\times \left(\left[-(n + \lambda + \frac{3}{4})x + \frac{\gamma}{2}\left(\frac{2n - 3m - \lambda + 3}{-n + 2m - 2}\right)\right](x^2 + 1)^{-\frac{1}{2}}\right.$$
$$\left. + \frac{\gamma}{2}(1 - x)^{-1}(x^2 + 1)^{\frac{1}{2}}\psi^{(2)}_{n,m}(x) - \left[\frac{i(-n + 2m + \lambda - 2)^2 + \frac{\gamma^2}{8}}{-n + 2m + \lambda + -2}\right](x^2 + 1)^{-\frac{1}{2}}\psi^{(2)}_{n-1,m}(x)\right) (3.19)$$

The wave functions $\psi_{n,m}(x)$ can be also satisfied in the second-order differential equations for each potential, correlatively,

$$(x^2 - 1)\ddot{\psi}^{(1)}_{n,m}(x) + \left[\frac{1}{2}(1 - \gamma - \lambda)x^3 + \frac{1}{2}(\gamma - \lambda)x^2 - 1\right]\dot{\psi}^{(1)}_{n,m}(x) +$$
$$\left[(1 - m)(\lambda + \gamma + m)x^2 + (m - 2)(\gamma - \lambda)x + \frac{(\lambda + \gamma + n + m)(m - n - 1)}{x^2 - 1}\right.$$
$$\left. + \frac{(\lambda + \gamma + 2m - 1)^2}{2} - m + 1\right]\psi^{(1)}_{n,m}(x) = 0. \quad (3.20)$$

and

$$(x^2 + 1)\ddot{\psi}^{(2)}_{n,m}(x) + [2(\lambda + 1)x + \gamma]\dot{\psi}^{(2)}_{n,m}(x) +$$

$$\left[(m^2 - n^2 - n) + 2\lambda(m - n) - \frac{(m^2 + 2m\lambda)x^2 - \gamma m x}{x^2 + 1}\right]\psi_{n,m}^{(2)}(x) = 0 \quad (3.21)$$

Rodrigues representations of the associated polynomials $\psi_{n,m}(x)$ are given by

$$\psi_{n,m}^{(1)}(x) = (-1)^m (x^2 - 1)^{\frac{m}{2}} \left(\frac{d}{dx}\right)^m \psi_n^{(1)}(x), \quad m = 0, 1, 2, .., n \quad (3.22)$$

and

$$\psi_{n,m}^{(2)}(x) = (-1)^m (x^2 + 1)^{\frac{m}{2}} \left(\frac{d}{dx}\right)^m \psi_n^{(2)}(x), \quad m = 0, 1, 2, .., n \quad (3.23)$$

where $\psi_n(x)$ satisfies in Jacobi differential equation which their Rodrigues representations, respectively, are

$$\psi_n^{(1)}(x) = N(x-1)^{-\lambda}(x+1)^{-\gamma}\left(\frac{d}{dx}\right)^n\left((x-1)^{n+\lambda}(x+1)^{n+\gamma}\right), \quad (3.24)$$

and

$$\psi_n^{(2)}(x) = N(x^2 + 1)^{-\lambda}\exp(-\gamma \tan^{-1} x)\left(\frac{d}{dx}\right)^n\left((x^2 + 1)^{n+\lambda}\exp(-\gamma \tan^{-1} x)\right), \quad (3.25)$$

where $N$ is a normalization constant. Since the wave functions $\Theta_{n,m}(x)$ connected to the lower spinor $\theta_{n,m}(x)$ that have been calculated based on upper spinors $\varphi_{n,m}(x)$ and Rodriguies representations of upper spinors can be generalized to the wave function $\Theta_{n,m}(x)$ as $\psi_{n,m}(x)$ and $\psi_{n-1,m}(x)$. Therefore, the above Rodriguies representation can also be related to the lower spinor components.

## 4 Association of Rodruiges representation with Generalized Lauegrre polynomials

When Mörse potential $\omega^{(1)}(r) = -\frac{\gamma}{2}e^{-r} - m - \frac{\lambda}{2} + \frac{1}{2} - \frac{k}{r}$ and 3D-dimensional oscillator potential $\omega^{(2)}(r) = \frac{\gamma}{4}r - (\lambda + m - \frac{1}{2})\frac{2}{r} - \frac{k}{r}$ are considered as the gauge field potentials, the upper spinor components are associated with Generalized Lauegrre polynomials. So, the second-order differential equations for the upper spinor components are written according to Eq.(2.6)

$$-\frac{d^2\varphi_{n,m}^{(1)}(r)}{dr^2} + \left[\frac{\gamma^2}{4}e^{-2r} + \gamma(m + \frac{\lambda}{2} - 1)e^{-r}\right]\varphi_{n,m}^{(1)}(r) = \left[\frac{(\alpha^2\eta - \varepsilon)^2 - 1}{\alpha^2}\right]\varphi_{n,m}^{(1)}(r), \quad (4.1)$$

$$-\frac{d^2\varphi_{n,m}^{(2)}(r)}{dr^2} + \left[\frac{\gamma^2}{16}r^2 + (\lambda + m - \frac{1}{2})(\lambda + m - \frac{3}{2})\frac{1}{r^2}\frac{\gamma}{2}(\lambda + m)\right]\varphi_{n,m}^{(2)}(r) = \left[\frac{(\alpha^2\eta - \varepsilon)^{(2)} - 1}{\alpha^2}\right]\varphi_{n,m}^{(2)}(r) (4.2)$$

According to the non-relativistic energy spectrum $E_{n,m}^{(1)} = -(n - m + 1)(\lambda + n + m)$ and $E_{n,m}^{(2)} = \gamma(n - m + 1)$, the relativistic energy spectrums are obtained as

$$\varepsilon_n^{(1)} = \alpha^2\eta \mp \left[-\alpha^2(n - m + 1)(\lambda + n + m) + 1\right]^{\frac{1}{2}}, \quad (4.3)$$

$$\varepsilon_n^{(2)} = \alpha^2\eta \mp \left[-\alpha^2\gamma(n - m + 1) + 1\right]^{\frac{1}{2}}. \quad (4.4)$$

The second-order differential equations(4.1) and (4.2) due to the solutions based on the Generalized Laugerre polynomials as upper spinor wave functions

$$\varphi_{n,m}^{(1)}(x) \propto \left(\frac{\gamma}{x}\right)^{-(n+\frac{\lambda}{2}+\frac{1}{2})} \times \exp\left(-\frac{\gamma}{2x}\right) L_n^{(-2n-\lambda-1)}\left(\frac{\gamma}{x}\right), \quad (4.5)$$

$$\varphi_{n,m}^{(2)}(x) \propto (\gamma x)^{\frac{\lambda+m-\frac{1}{2}}{2}} \times \exp\left(-\frac{\gamma x}{2}\right) L_n^{(\lambda+m-\frac{1}{4})}(\gamma x), \quad (4.6)$$

where $L_n^\alpha(x)$ is Generalized Laugerre polynomial with $\alpha > -1$. In the upper spinor (4.5), $x = e^r$ and $\alpha = -2n - \lambda - 1$ and in the other upper spinor (4.6), $x = \frac{r^2}{4}$ and $\alpha = \lambda + m - \frac{1}{4}$. Lower spinor components can be attained Eq.(2.7) for each potential

$$\theta_{n,m}^{(1)}(x) \propto \left(\frac{1}{\alpha} \mp \left[\frac{1}{\alpha^2} + (n - m + 1)(-m - n - \lambda)\right]^{\frac{1}{2}}\right)^{-1}$$
$$\times \left(\left[\frac{(-2m - \lambda + 1) - (n - m + 1)}{\gamma}\right] x \, \varphi_{n-1,m}^{(1)}(x), \quad (4.7)\right.$$

$$\theta_{n,m}^{(2)}(x) \propto \left(\frac{1}{\alpha} \mp \left[\frac{1}{\alpha^2} + \gamma(n-m+1)\right]^{\frac{1}{2}}\right)^{-1}$$
$$\times x^{-\frac{1}{2}} \left(\left[(n-m+1)\varphi_{n,m}^{(2)}(\gamma x) - (n+\lambda+\frac{3}{4})\varphi_{n-1,m}^{(2)}(\gamma x)\right]\right) \qquad (4.8)$$

As mentioned in pervious section, the raising and lowering $B_+$ and $B_-$ operators based on the superpotentials are given as

$$B_+^{(1)} = \frac{d}{dr} - \frac{\gamma}{2}e^{-r} - m - \frac{\lambda}{2} + \frac{1}{2}, \qquad B_-^{(1)} = -\frac{d}{dr} - \frac{\gamma}{2}e^{-r} - m - \frac{\lambda}{2} + \frac{1}{2}, \qquad (4.9)$$

$$B_+^{(2)} = \frac{d}{dr} + \frac{\gamma}{4}r - (\lambda + m - \frac{1}{2})\frac{2}{r}, \qquad B_-^{(2)} = -\frac{d}{dr} + \frac{\gamma}{4}r - (\lambda + m - \frac{1}{2})\frac{2}{r}. \qquad (4.10)$$

The pair of linear differential operators can factorize Schrödinger equation for each potential. Similar to the pervious section, in Mörse potential, if $A^{(1)}(x) = x^2$ and $W^{(1)}(x) = x^\lambda e^{-\frac{\gamma}{x}}$, the wave function $\psi_{n,m}^{(1)}(x)$ is written based on upper spinor as

$$\psi_{n,m}^{(1)}(x) = \left(\frac{\exp(\frac{\gamma}{2x})}{x}\right)\varphi_{n,m}^{(1)}(x), \qquad (4.11)$$

and, in 3D-dimensional oscillator potential, when $A^{(2)}(x) = x$ and $W^{(2)}(x) = x^\lambda e^{-\gamma x}$, the wave function $\psi_{n,m}^{(2)}(x)$ is obtained as:

$$\psi_{n,m}^{(2)}(x) = x^{-(\frac{\lambda}{2}+\frac{1}{4})} \exp(\frac{\gamma x}{2})\varphi_{n,m}^{(2)}(x). \qquad (4.12)$$

Therefore, they are also related to the Generalized Laugerre polynomials. It is clear that, both of them are examples of associating Generalized Laugerre polynomial with Rodriguies representation in Dirac equation. Also, Rodriguies representation of lower spinors $\theta_{n,m}(x)$ tht are called $\Theta_{n,m}(x)$ will be the following forms based on $\psi_{n,m}(x)$ and $\psi_{n-1,m}(x)$, for Mörse and 3D-dimensional oscillator potentials

$$\Theta_{n,m}^{(1)}(x) \propto \left(\frac{1}{\alpha} \mp \left[\frac{1}{\alpha^2} + (n-m+1)(-m-n-\lambda)\right]^{\frac{1}{2}}\right)^{-1}$$
$$\times x \left[-\frac{(m+n+\lambda)}{\gamma}\right]\psi_{n-1,m}^{(1)}(x), \qquad (4.13)$$

$$\Theta^{(2)}_{n,m}(x) \propto \left( \frac{1}{\alpha} \mp \left[ \frac{1}{\alpha^2} + \gamma(n-m+1) \right]^{\frac{1}{2}} \right)^{-1}$$
$$\times x^{\frac{\lambda+1}{2}} \left[ (n-m+1)\psi^{(2)}_{n,m}(x) + (n+\lambda+\frac{3}{4})\psi^{(2)}_{n-1,m}(x) \right]. \quad (4.14)$$

The wave functions $\psi_{n,m}(x)$ can be also satisfied in second-order differential equations for each potential, proportionately,

$$x^2 \ddot{\psi}^{(1)}_{n,m}(x) + (\lambda x + \gamma + 2)\dot{\psi}^{(1)}_{n,m}(x) + \left( -\frac{m\gamma}{x} + m^2 - m - 2n^2 - n(\lambda+1) \right) \psi^{(1)}_{n,m}(x) = 0, \quad (4.15)$$

and

$$x\ddot{\psi}^{(2)}_{n,m}(x) + \left( 2(\lambda + \frac{1}{2}) - \gamma x \right) \dot{\psi}^{(2)}_{n,m}(x) +$$
$$\left( \gamma(n - m + \frac{1}{2}) - \frac{\gamma\lambda}{2} - \frac{(m-1)(2\lambda+1)}{4} x^{-\frac{1}{2}} \right) \psi^{(2)}_{n,m}(x) = 0. \quad (4.16)$$

Rodrigues representations of associated polynomials $\psi_{n,m}(x)$ are given as

$$\psi^{(1)}_{n,m}(x) = (-1)^m x^m \left( \frac{d}{dx} \right)^m \psi^{(1)}_n(x), \quad m = 0, 1, 2, .., n \quad (4.17)$$

and

$$\psi^{(2)}_{n,m}(x) = (-1)^m x^{\frac{m}{2}} \left( \frac{d}{dx} \right)^m \psi^{(2)}_n(x), \quad m = 0, 1, 2, .., n \quad (4.18)$$

where $\psi_n(x)$ satisfies in Laugerre differential equation which their Rodrigues representations are, respectively,

$$\psi^{(1)}_n(x) = N x^{-\lambda} \exp(\frac{\gamma}{x}) \left( \frac{d}{dx} \right)^n \left( x^{2n+\lambda} \exp(-\frac{\gamma}{x}) \right), \quad (4.19)$$

and

$$\psi_n^{(2)}(x) = Nx^{-\lambda}\exp(\gamma x)\left(\frac{d}{dx}\right)^n\left(x^{\lambda+n}\exp(-\gamma x)\right). \tag{4.20}$$

The above Rodrigues representations can be also related to lower spinor components, because of there are the wave function $\Theta_{n,m}(x)$ based on the wave function $\psi_{n,m}(x)$ and $\psi_{n-1,m}(x)$ according to the lower spinor components.

## 5  Association of Rodruiges representation with Hermite polynomials

The upper spinor component will be considered as Hermite polynomials, if the gauge field potential $\omega(r) = \frac{\gamma}{2}r - \lambda - \frac{k}{r}$ is written based on shift-oscillator potential. This upper spinor component satisfies in Eq.(2.6) as:

$$-\frac{d^2\varphi_{n,m}(r)}{dr^2} + \left[\left(\frac{\gamma}{2}r - \lambda\right)^2 - \frac{\gamma}{2}\right]\varphi_{n,m}(r) = \left[\frac{(\alpha^2\eta - \varepsilon)^2 - 1}{\alpha^2}\right]\varphi_{n,m}(r). \tag{5.1}$$

For this potential, the non-relativistic energy $E_{n,m} = \gamma(n - m + 1)$ can be used to the following relativistic energy spectrum

$$\varepsilon_n = \alpha^2\eta \mp \left[\alpha^2\gamma(n - m + 1) + 1\right]^{\frac{1}{2}}. \tag{5.2}$$

The upper spinor wave function based on the Hermite polynomials can be obtained from Eq.(5.1)

$$\varphi_{n,m}(x) \propto exp\left(-\frac{\gamma}{4}x^2\right) \times H_n\left(\left(\frac{\gamma}{2}\right)^{\frac{1}{2}}x\right), \tag{5.3}$$

where $H_n(x)$ is Hermite polynomial. In the wave function (5.3) $x = r - \frac{2\lambda}{\gamma}$ and $-\infty < x < +\infty$. According to Eq.(2.7), the lower spinor wave function is calculated as

$$\theta_{n,m}(x) \propto \left(\frac{1}{\alpha} \mp \left[\frac{1}{\alpha^2} + \gamma(n - m + 1)\right]^{\frac{1}{2}}\right)^{-1}$$

$$\times \left( \left[ (2\gamma)^{\frac{1}{2}} (n-m+1) \right] \varphi_{n-1,m}(x). \right. \tag{5.4}$$

To factorize Schrödinger equation, there are pairs of linear differential operator as

$$B^+ = \frac{d}{dr} + \frac{\gamma}{2} r - \lambda, \qquad B^- = -\frac{d}{dr} + \frac{\gamma}{2} r - \lambda. \tag{5.5}$$

As mentioned before, the wave function $\psi_{n,m}(x)$ that is related to upper spinor is gotten by

$$\psi_{n,m}(x) = \exp(\frac{\gamma}{4} x^2) \varphi_{n,m}(x), \tag{5.6}$$

for $A(x) = 1$ and $W(x) = \exp\left(-\frac{\gamma}{2} x^2\right)$. In this potential, Hermite polynomial can be associated with Rodrigues representation in Dirac equation. Also, similar to pervious section, the wave function $\Theta_{n,m}(x)$ that is connected to lower spinor $\theta_{n.m}(x)$ can be written based on Rodrigues representation

$$\Theta_{n,m}(x) \propto \left( \frac{1}{\alpha} \mp \left[ \frac{1}{\alpha^2} + \gamma(n-m+1) \right]^{\frac{1}{2}} \right)^{-1}$$
$$\times \left[ (2\gamma)^{\frac{1}{2}} (n-m+1) \right] \psi_{n-1,m}(x). \tag{5.7}$$

The second-order differential equation for shift-oscillator potential will be

$$\ddot{\psi}_{n,m}(x) + (-\gamma x) \dot{\psi}_{n,m}(x) - \gamma(m-n) \psi_{n,m}(x) = 0, \tag{5.8}$$

where Rodrigues representation of the associated polynomial $\psi_{n,m}(x)$ is considered by

$$\psi_{n,m}(x) = (-1)^m \left( \frac{d}{dx} \right)^m \psi_n(x), \quad m = 0, 1, 2, ..., n \tag{5.9}$$

and Rodrigues representation of $\psi_n(x)$, will be the following form

$$\psi_n(x) = N \exp(\frac{\gamma}{2} x^2) \left( \frac{d}{dx} \right)^n \left( \exp(-\frac{\gamma}{2} x^2) \right). \tag{5.10}$$

Since $\Theta_{n,m}(x)$ has been used for lower spinor $\Theta_{n,m}(x)$ and it has been associated with $\psi_{n-1,m}(x)$ according (5.7), therefore, the Eqs.(5.8), (5.9) and (5.10) can be considered for the lower spinor component in shift-oscillator potential.

## 6 Conclusion

It has presented a procedure for connecting the methods used in the analysis of exactly solvable potentials in the non-relativistic quantum mechanics with the solution of Dirac equation. A gauge field potential and the bound states energy spectrum have been defined for the Dirac equation with a constant electrostatic potential that can be constituted a Schrödinger-like equation. Since orthogonal polynomials are considered as the solution of Schrödinger-like equation that have been obtained from Dirac equation, Rodrigues representations of the orthogonal polynomials can be associated with upper and lower spinor components.